\documentclass{optica-article}
\usepackage[utf8]{inputenc}
\usepackage[T1]{fontenc}
\usepackage{physics}
\usepackage{float}
\usepackage{cleveref}
\usepackage{mathtools}
\usepackage{pythonhighlight}
\usepackage{bm}
\usepackage{amsmath}
\usepackage{booktabs}
\usepackage{siunitx}
\usepackage{float}
\usepackage{dsfont}
\usepackage{cases}

\journal{opticajournal} 

\articletype{Research Article}

\usepackage{lineno}
\linenumbers 

\begin{document}
\nolinenumbers

\title{Analyzing the acceleration time and reflectance of light sails made from homogeneous and core-shell spheres}

\author{Mitchell R. Whittam,\authormark{1,*} Lukas Rebholz,\authormark{1} Benedikt Zerulla,\authormark{2} Carsten Rockstuhl\authormark{1, 2}}

\address{\authormark{1}Institute of Theoretical Solid State Physics, Karlsruhe Institute of Technology (KIT), Kaiserstr. 12, 76131 Karlsruhe, Germany\\
\authormark{2}Institute of Nanotechnology, Karlsruhe Institute of Technology (KIT), Kaiserstr. 12, 76131 Karlsruhe, Germany}

\email{\authormark{*}mitchell.whittam@kit.edu} 



\begin{abstract*}
Deciding on appropriate materials and designs for use in light sails, like the one proposed in the Breakthrough Starshot Initiative, is a topic that requires much care and forethought. Here, we offer a feasible option in the form of metasurfaces made of periodically arranged homogeneous and core-shell spheres. Using the re-normalized T--matrix from Mie theory, we explore the reflectance, absorptance, and acceleration time of such metasurfaces. We focus on spheres made from aluminum, silicon, silicon dioxide, and combinations thereof. Since the light sails are foreseen to be accelerated using Earth-based laser arrays to 20\% of the speed of light, one needs to account for relativistic effects. As a result, a high broadband reflectance is essential for effective propulsion. We identify metasurfaces that offer such properties combined with a low absorptance to reduce heating and deformation. We highlight a promising extension to the case of a metasurface made from homogeneous silicon spheres, as already discussed in the literature, by adding a layer of silicon dioxide. The high broadband reflectance of the silicon and silicon dioxide combination is explained by the favorable interference of the multipolar contributions of the outgoing field up to quadrupolar order. We also consider the impact of an embedding material characterized by different refractive indices. Refractive indices up to 1.13 maintain over 90\% reflectance without re-optimizing the light sail.
\end{abstract*}


\flushbottom
%
%
\section*{Introduction}
The Breakthrough Starshot Initiative was announced in 2016 with the aim of propelling gram-scale microsatellites into outer space for use in rapid solar system missions. The microsatellites are to be equipped with a light sail and accelerated using Earth-based laser arrays to 20\% of the speed of light~\cite{parkin2018breakthrough, daukantas2017breakthrough}. Before this project can become a reality, many engineering challenges must be overcome, involving the thermal management of the light sail~\cite{brewer2022multiscale, jin2022laser, holdman2022thermal, tung2022low}, ensuring its stability in the acceleration process~\cite{rafat2022self, savu2022structural, gieseler2021self, manchester2017stability}, and accounting for different forces acting on the sail~\cite{achouri2017metasurface}. Another critical aspect concerns the materials from which the light sail should be made. Many materials have been suggested recently~\cite{atwater2018materials, campbell2021relativistic}, along with potential ways to structure and combine them. Examples include single and multilayered slabs~\cite{santi2022multilayers, jin2020inverse, myilswamy2020photonic, chang2024broadband}. It was also suggested to exploit metasurfaces~\cite{schulz2024roadmap, choudhury2018material}, specifically those made from periodically arranged silicon spheres~\cite{evlyukhin2020lightweight}. 

In our work, we build upon the latter but specifically consider metasurface light sails made from core-(multi-)shell spheres arranged in a lattice structure. We compare their performance to light sails containing homogeneous spheres, i.e., spheres made from a single material. Our theoretical and numerical work relies on the T--matrix formalism in the context of Mie theory to describe the scattering from the spheres~\cite{bohren2008absorption, mishchenko2002scattering}. We fully account for lattice effects~\cite{babicheva2021multipole} to compute the reflectance, transmittance, and absorptance from these metasurfaces by considering the re-normalized T--matrix, where the T--matrix of an isolated sphere is re-normalized by the interaction with all the other particles in the lattice~\cite{Beutel:21}. Initially, we assume that the spheres are suspended in a vacuum, but we discuss including an embedding material in a later chapter. The specific materials we consider are silicon (Si), silicon dioxide (SiO$_{2}$), and aluminum (Al).

When selecting the possible materials, there are many criteria one must consider. Firstly, the materials should exhibit a high reflectance across the entire acceleration duration to ensure maximum momentum transfer from the lasers. To do this, one must consider the effects of Doppler shifts due to the relativistic motion of the sail. In other words, as the sail moves away from the light source, it will observe an incident field with a longer wavelength $\lambda'$ compared to the wavelength $\lambda_{0}$ as viewed by a stationary observer on Earth. Quantitatively, during the acceleration process, the sail will observe wavelengths in the range of $\lambda'\in[1.0\lambda_{0}, 1.225\lambda_{0}]$, meaning the reflectance should be as large as possible for each of these wavelengths. 

Secondly, the absorptance of the materials should be kept to a minimum to limit unwanted temperature increases in the sail. Avoiding the generation of excess heat is important since the sail could otherwise experience material deformations such as melting. Possible ways to mitigate this include coating the spheres with a highly reflective outer layer (e.g., aluminum) to deflect as much heat as possible away from the sail. Another method would be to use materials with a low absorptance in the desired wavelength range (e.g., Si and SiO$_{2}$). Moreover, radiative cooling is an essential factor to consider, a process that measures how well an object can lose heat by thermal radiation~\cite{zhao2019radiative, fan2022photonics, zhou2019polydimethylsiloxane}. An example of how to account for radiative cooling was investigated in Refs.~\cite{holdman2022thermal, ilic2018nanophotonic}, where it is proposed to add SiO$_{2}$ to a layer of Si. We extend this by highlighting a combination of Si and SiO$_{2}$ in core-shell spheres.  

Finally, the sail must have a mass as low as possible to accelerate quickly. To incorporate this aspect, as well as the reflectance and absorptance of the sail, one must consider a comprehensive figure of merit. Quantities discussed in the past include the acceleration distance (the distance required to reach the target speed of 20\% of the speed of light)~\cite{mcinnes2004solar, macchi2009light, atwater2018materials, kudyshev2021optimizing} and the acceleration time (the time required to reach the target speed)~\cite{santi2022multilayers}. In our contribution, we consider the latter, intending to keep the acceleration time as low as possible. For completion, we also include data relevant to a numerically minimized acceleration time that makes no prior assumption about the reflectance and absorptance. We also explain why it might be preferred to determine the acceleration time after first obtaining a high broadband reflectance and low broadband absorptance. 

The structure of the paper is as follows: in the opening section, we outline the setup and quantitatively define the acceleration time. We also mathematically express the re-normalized T--matrix of a lattice in terms of the T--matrix of an isolated sphere, the quantity crucial to study numerically and analytically the optical response of the light sail. Afterward, the reflectance, absorptance, and acceleration times of lattices containing spheres made from one material will be examined for each of the abovementioned materials. Following this, we analyze lattices containing core-(multi-)shell spheres, highlighting the combination of Si and SiO$_{2}$ due to the corresponding high broadband reflectance. Moreover, we provide physical insights into the high broadband reflectance by analyzing the contribution of each multipolar order to the outgoing field up to quadrupolar order. Thanks to our formalism, we can also explore how lattice and multipolar interactions~\cite{kiselev2020multipole} promote the preferable response. Next, we discuss the possible inclusion and effects of an embedding for the spheres. Finally, we conclude our findings.

\section*{Description of the Scattering Scenario}
\begin{figure}[h!]
\centering
\includegraphics[width=1.0\textwidth]{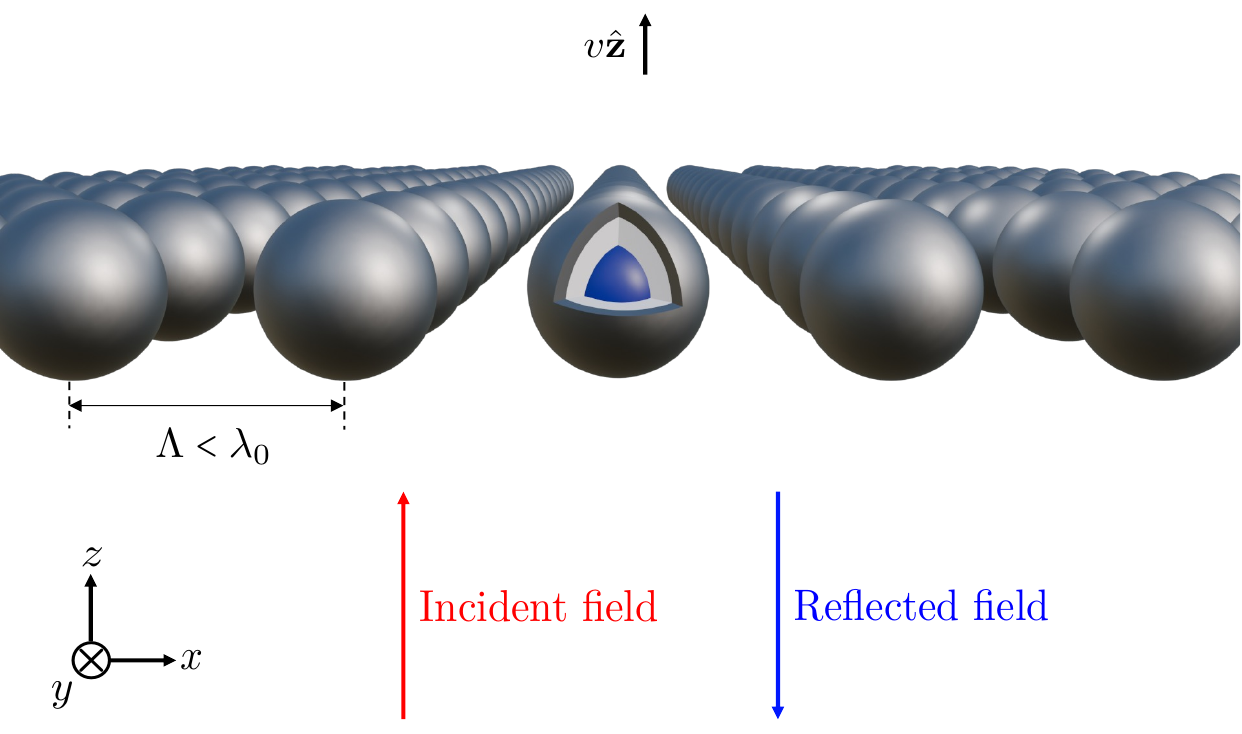}
\caption{A light sail infinitely spanning the $x$-$y$ plane moving along the $+z$-axis at speed $v$ illuminated by a linearly polarized incident plane wave reflected in the $-z$ direction. The sail comprises spheres (which we depict here as having a core and two shells) arranged in a square lattice characterized by a lattice constant $\Lambda$, which is always smaller than the wavelength $\lambda_{0}$ of the incident field as observed on Earth.}
\label{fig:metasurface}
\end{figure}

We begin by outlining the setup in our investigation. We consider a light sail comprising periodically arranged spheres infinitely spanning the $x$-$y$ plane while moving with speed $v$ along the $+z$-axis (c.f. Fig~\ref{fig:metasurface}). Initially, the spheres are homogeneous and made from a single material. Afterward, we also consider spheres made from a core and one or two shells of different materials. At this stage, we omit the inclusion of a substrate and concentrate solely on optimizing the metasurface and its constituents. We discuss in a later chapter the inclusion of a supporting embedding. Our model assumes a linearly polarized plane wave incident field illuminating the metasurface characterized by a square lattice with lattice constant $\Lambda$ and unit cell area $\Lambda^{2}$. The source of the incident field is an Earth-based laser with intensity $I$ that propels the sail.

As quoted in the literature~\cite{rafat2022self, salary2020photonic}, a range of possible wavelengths $\lambda_{0}$ of the incident field as observed on Earth is $\SI{1.0}{\micro\meter}-\SI{1.5}{\micro\meter}$. For our purposes, we consider a wavelength of $\lambda_{0}=\SI{1.0}{\micro\meter}$, such that our results cover a broadband spectrum of $\lambda' \in [1.0\lambda_{0}, 1.225\lambda_{0}]$ in the scatterer's frame due to the Doppler shift defined as
\begin{align}\label{eq:doppler_shift}
    \lambda'(\beta) = \lambda_{0}(1+\beta)\gamma(\beta)\, .
\end{align}
Here, $\beta=v/c$, $c$ is the speed of light in vacuum, and $\gamma(\beta)=1/\sqrt{1-\beta^{2}}$ is the Lorentz factor. 
We emphasize that we only consider sub-wavelength lattice constants and sphere radii to eliminate higher diffraction orders. In other words, we only consider the zeroth diffraction order such that reflection solely occurs along the $-z$-axis.   

In our scenario, we consider a light sail that is accelerated to a final target speed of $v_{\mathrm{f}}=\beta_{\mathrm{f}}c$, where $\beta_{\mathrm{f}}=0.2$. Such a restriction aligns with assumptions in the context of the Breakthrough Starshot Initiative~\cite{parkin2018breakthrough, daukantas2017breakthrough}.

The quantities we wish to analyze are the reflectance $R(\beta)$, absorptance $A(\beta)$, and the acceleration time $\tau$ as defined by Eqn.~(1) in~\cite{santi2022multilayers}. This acceleration time is the time required to accelerate the sail to the desired target speed $v_{\mathrm{f}}$. Specifically, we will explore with these properties the effectiveness of various combinations of materials for use in a light sail.

Since we consider an infinitely large metasurface, we express $\tau$ in terms of densities instead of masses as follows:
\begin{align}\label{eq:acc_time}
    \tau
    &=
    \frac{c^{2}}{I\Lambda^{2}}\int^{\beta_{\mathrm{f}}}_{0}\frac{\mu_{\mathrm{p}}+\sum_{j}\rho_{j}V_{j}}{A(\beta)+2R(\beta)}\frac{\gamma^{3}(1+\beta)}{(1-\beta)}\mathrm{d}\beta\, ,
\end{align}
where $R(\beta)$ and $A(\beta)$ are the reflectance and absorptance as a function of the speed ratio $\beta$, respectively. Note that each value for $\beta$ corresponds to a unique wavelength in the aforementioned range $\lambda'\in[1.0\lambda_{0}, 1.225\lambda_{0}]$, each of which is determined from Eq.~\eqref{eq:doppler_shift}. Moreover,  $\mu_{\mathrm{p}}=m_{\mathrm{p}}\Lambda^{2}/a$ is the mass of the payload per unit cell for a given sail of area $a$, and $\rho_{j}$ and $V_{j}$ are the density and volume of the $j$'th constituent material of each sphere, respectively. One should keep in mind that the lattice constant remains invariant despite the relativistic motion, as $\Lambda$ is only defined in the $x$-$y$ plane where no length contraction occurs due to the motion solely occurring along the $z$-axis. Moreover, coupled with the sail always observing a wavelength $\lambda'\geq\lambda_{0}$, we can be confident that the system always remains subwavelength.

We emphasize that we consider an acceleration time $\tau$ defined in Eq.~\eqref{eq:acc_time} that does not explicitly depend on temperature. Furthermore, we assume room-temperature values for all relevant material parameters. However, these two assumptions suffice for our purposes since we aim to demonstrate an optimization method one can use for a given set of parameters. One could, of course, implement the same method for temperature-dependent quantities, but this would require identifying new optimized parameters.

In our investigation, we firstly minimize the acceleration time $\tau$ as much as possible numerically concerning all parameters in Eq.~\eqref{eq:acc_time} without making any assumptions about the reflectance $R(\beta)$ or absorptance $A(\beta)$. Then, we present results where the reflectance is as large as possible and the absorptance is correspondingly as small as possible across all wavelengths. Why one might wish to ensure simultaneously a high broadband reflectance and low broadband absorptance can be justified as follows: firstly, a high reflectance means a high momentum transfer to the light sail, thus propelling it more quickly. Secondly, the acceleration time $\tau$ defined in Eq.~\eqref{eq:acc_time} does not take into account temperature changes of the sail, which would appear if the absorptance is high. As we will see for spheres containing a SiO$_{2}$ core, central Si shell, and outer Al shell, the acceleration time is very low, but the average broadband absorptance is high (c.f. Table~\ref{table:table_tau_r_a}). A high broadband absorptance is unfavorable and could lead to thermal deformation in the sail. 

Since the acceleration time depends on the reflectance $R(\beta)$ and absorptance $A(\beta)$ of the sail, we need to obtain the outgoing field from the considered light sail. The process for determining the outgoing field from a lattice of spheres relies on two matrices, specifically the re-normalized T--matrix $\mathbf{T}_{\mathrm{ren}}$ and the S--matrix $\mathbf{S}$. The re-normalized T--matrix describes the scattering of one scatterer at the origin of a two-dimensional lattice, incorporating the lattice interaction. It is defined as 
\begin{align}\label{eq:MultiScat}
    \mathbf{T}_{\mathrm{ren}}=\left(\mathds{1}-\mathbf{T}\sum_{\bm{R}\neq 0}\mathbf{C}^{(3)}(-\bm{R})\mathrm{e}^{\mathrm{i}\bm{k}_{\parallel}(\omega)\bm{R}}\right)^{-1}\mathbf{T}\ .
\end{align}
See also Eq.~(17) from Ref.~\cite{Beutel:21}. In Eq.~(\ref{eq:MultiScat}), $\mathbf{T}$ is the T--matrix of the individual scatterer, $\bm{R}$ is a two-dimensional lattice vector, $\mathbf{C}^{(3)}(-\bm{R})$ is a translation matrix, and $\bm{k}_{\parallel}$ is the component of the wave vector of the incident plane wave with frequency $\omega$ parallel to the lattice. The re-normalized T--matrix, also called the effective T--matrix in Ref.~\cite{https://doi.org/10.1002/adom.202102059}, gives information about the optical response of a scatterer in a two-dimensional lattice. With the re-normalized T--matrix at hand, one can compute the scattered electric field of the lattice (c.f. Eqs.~(21,22) from Ref.~\cite{Beutel:21}) and finally the S--matrix $\mathbf{S}$, relating the incoming field to the outgoing field (c.f. Eq.~(6) from Ref.~\cite{Beutel:21}). Note that in Ref.~\cite{Beutel:21}, the S--matrix is referred to as the Q--matrix.

\section*{Individual materials}
Before investigating the acceleration time of a light sail comprising core-shell spheres, we first explore the properties of each constituent material separately.  The specific materials we consider are silicon (Si), silicon dioxide (SiO$_{2}$), and aluminum (Al). These materials have already been considered in works related to the Breakthrough Starshot Initiative~\cite{santi2022multilayers, rajalakshmi2017fuel}. Aluminum is a good reflector as a metal, which is important to ensure maximum momentum transfer to the sail. Moreover, the remaining materials are nearly non-absorbing. A vanishing absorption is vital, as the temperature increase of the sail must be kept to a minimum to avoid material deformations like melting. 

Crucially, the densities of each material (taken from Ref~\cite{haynes2014crc} and given in Table~\ref{table:densities_combinations}) are low, which reduces the necessary acceleration time.
\begin{table}
\bigskip
\centering
\caption{The densities of Si, SiO$_{2}$ and Al~\cite{haynes2014crc}.}
\begin{tabular}{r|rrrr}
\toprule
Material & SiO$_{2}$ & Si & Al \\
\midrule
Density (kg~m$^{-2}$) & $2196$ & $2330$ & $2700$\\
\bottomrule
\end{tabular}
\label{table:densities_combinations}
\end{table}

Beginning with determining the reflectance $R(\beta)$ and absorptance $A(\beta)$ with the help of Eq.~\eqref{eq:doppler_shift}, we exploit the software package \emph{treams}~\cite{beutel2024treams} and perform a parameter sweep across $N_{\beta}=50$ equally spaced $\beta$ values, $N_{\Lambda}=120$ equidistant values of the lattice constant $\Lambda$, and $N_{r}=110$ equidistant values for the radius $r$ of each sphere. We ensure that $r<\Lambda/2$ and $\Lambda$ always remain subwavelength to avoid additional diffraction orders. We implement a parameter sweep instead of a more complicated optimization procedure since \emph{treams} already performs quickly. To ensure their convergence, the reflectance $R(\beta)$ and absorptance $A(\beta)$ are determined using entries from $\mathbf{T}_{\mathrm{ren}}$ up to hexadecapolar (fourth) order. However, when we later consider metasurfaces consisting Si/SiO$_{2}$ spheres, convergence already occurs at quadrupolar order.    

As previously mentioned, we consider two sets of data. Firstly, we determine the numerically minimized acceleration time $\tau_{\mathrm{min}}$, where no previous assumptions about the reflectance $R(\beta)$ or absorptance $A(\beta)$ are made. Then, we calculate the acceleration time $\tau_{\mathrm{bb}}$ corresponding to a high broadband reflectance, where the `bb' subscript denotes `broadband'. To obtain a high broadband reflectance, we rely on maximizing the average reflectance $\overline{R}$, where 
\begin{align}\label{eq:ratio}
    \overline{R}
    &=
    \frac{1}{N_{\beta}}\sum_{j=1}^{N_{\beta}}R(\beta_{j})\, .
\end{align}
The reason for considering the average reflection is that it allows us to locate points where the reflectance could be uniformly high across all wavelengths as long as the absorptance remains low. A high reflectance is imperative to ensure maximum momentum transfer to the sail, and a low absorptance reduces unwanted heating of the light sail. 

\begin{figure}[h!]
\centering
\includegraphics[width=1\textwidth]{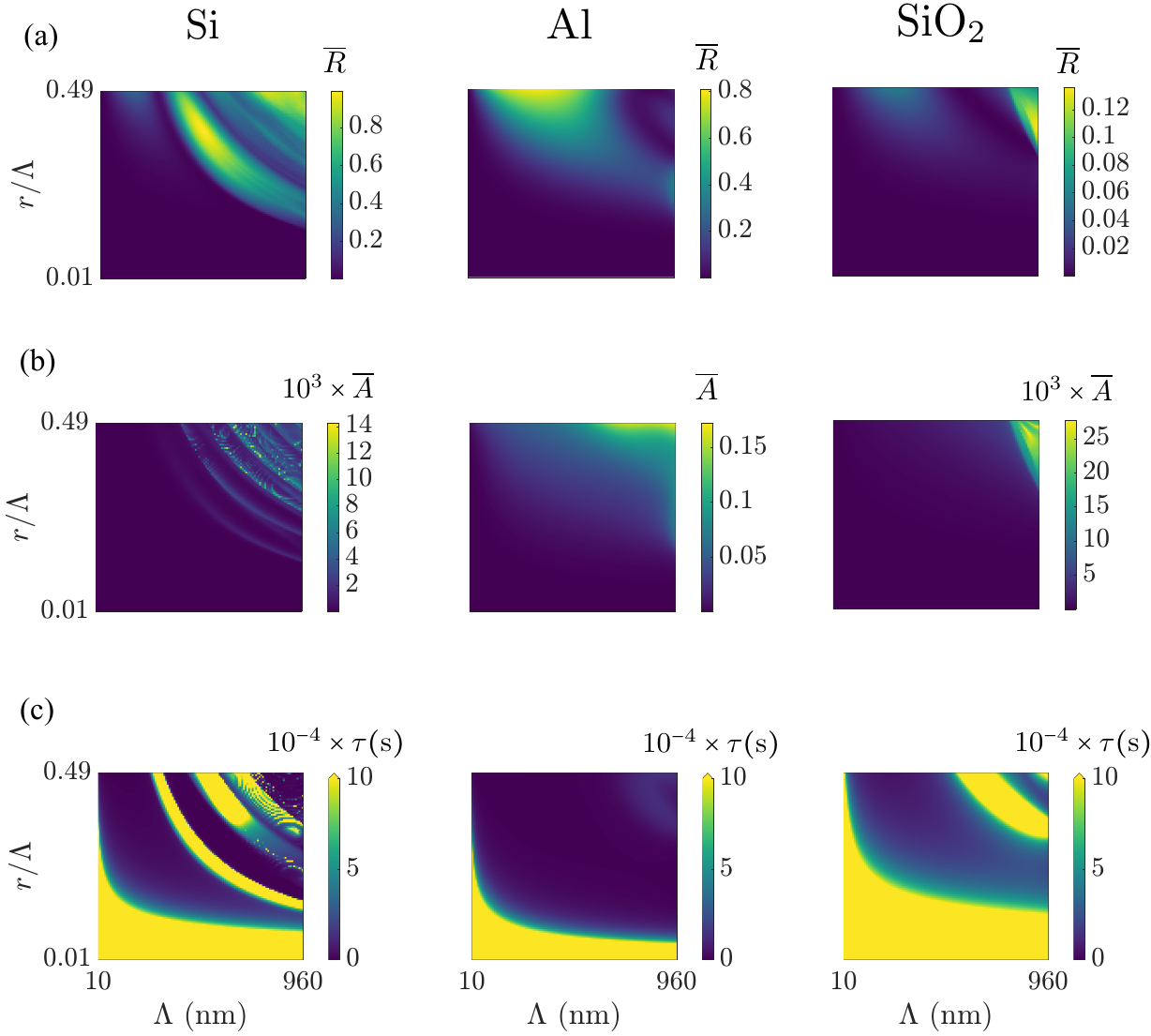}
\caption{In (a), the average reflectance $\overline{R}$ of lattices containing spheres made of silicon (Si), aluminum (Al), and silicon dioxide (SiO$_{2}$) as a function of the sphere radius $r$ and lattice constant $\Lambda$. In (b), the analogous plots are given for the average absorptance $\overline{A}$, while the corresponding acceleration times $\tau$ for each lattice are represented in (c).}
\label{fig:ref_avg}
\end{figure}

The considered refractive indices were based on experimental data from Refs.~\cite{schinke2015uncertainty, rodriguez2016self, mcpeak2015plasmonic} obtained from refractiveindex.info, which we interpolated to fit our desired wavelength range. The data for $\overline{R}$ and the average absorptance $\overline{A}$ for each material are given in Figs~\ref{fig:ref_avg}~(a) and~(b), respectively, along with the corresponding acceleration times $\tau$ in Fig~\ref{fig:ref_avg}~(c) calculated using Eq.~\eqref{eq:acc_time}. Note that, as in~\cite{atwater2018materials}, we use numerical values of $I=\SI{10}{\giga\watt\meter}^{-2}$, $m_{\mathrm{p}}=\SI{0.1}{\gram}$ and $a=\SI{10}{\meter}^{2}$. To make sure we remain in the subwavelength regime, we express the sphere radius $r$ in terms of the lattice constant $\Lambda$, where $0.01\leq\Lambda/\lambda_{0}\leq0.96$ and $0.01\leq r/\Lambda \leq0.49$.

A few features in Fig.~\ref{fig:ref_avg} are worth noting. Firstly, one finds that the array of Si spheres produces a very high maximum average reflectance of $\overline{R}_{\mathrm{max}}=0.989$. Moreover, a minimum acceleration time of $\tau_{\mathrm{min}}=\SI{212.7}{\second}$ occurs at $r=\SI{164.6}{\nano\meter}$ and $\Lambda=\SI{952.0}{\nano\meter}$. Conversely, SiO$_{2}$ exhibits a very poor maximum average reflectance of $\overline{R}_{\mathrm{max}}=0.136$. At the same time, Al has a high maximum average reflectance of $\overline{R}_{\mathrm{max}}=0.810$ but also an unfavorable corresponding maximum average absorptance of $\overline{A}_{\mathrm{max}}=0.058$ due to its high extinction coefficient. As a result, the SiO$_{2}$ and Al light sails have longer acceleration times than those made from Si spheres. In Table~\ref{table:table_tau_r_a_single}, we provide data for the numerically minimized acceleration time $\tau_{\mathrm{min}}$ and the acceleration time $\tau_{\mathrm{bb}}$ for the case of maximum average reflectance, as well as the relevant radii and lattice constants. 


As seen in Table~\ref{table:table_tau_r_a_single}, Si performs best out of all three materials, exhibiting the lowest minimized acceleration time $\tau_{\mathrm{min}}=\SI{212.7}{\second}$ and broadband acceleration time $\tau_{\mathrm{bb}}=\SI{326.7}{\second}$. Nevertheless, we once again emphasize that the acceleration time in Eq.~\eqref{eq:acc_time} is temperature independent and is based on room-temperature quantities. Consequently, although pure Si spheres yield favorable acceleration times, the corresponding light sail could fall foul to thermal damage. This is because the absorptance of Si increases quickly with temperature~\cite{holdman2022thermal}. To remedy this, it is suggested to add SiO$_{2}$ to the structure since the absorptance of SiO$_{2}$ exhibits minimal temperature variation in our desired wavelength range~\cite{wray1969refractive, holdman2022thermal}. The following sections will consider core-(multi-)shell structures, highlighting a combination containing Si and SiO$_{2}$ that produces a high broadband reflectance. Additionally, we will show how the multipolar interactions of the resulting lattice system combine to cause such a high reflectance.


\begin{table}[H]
\bigskip
\centering
\caption{Values for the numerically minimized acceleration times $\tau_{\mathrm{min}}$ without making any prior assumptions about the reflectance and absorptance, and those ($\tau_{\mathrm{bb}}$) that correspond to a maximum broadband reflectance, along with the relevant average reflectance $\overline{R}$, average absorptance $\overline{A}$, radii $r$, and lattice constants $\Lambda$ for each material.}
\begin{tabular}{rrrrrrrr}
\toprule
Materials & $r$ (nm) & $\Lambda$ (nm) &$\tau$ (s) & $\overline{R}$ & $\overline{A}$\\
\midrule
Si & $164.6$ & $952.0$ & $\tau_{\mathrm{min}}=212.7$ & $0.419$ & $4.2\times10^{-4}$\\
 & $187.0$ & $481.0$  & $\tau_{\mathrm{bb}}=326.7$ & $0.989$ & $4.7\times10^{-4}$\\
\midrule
Al & $40.1$ & $81.8$ & $\tau_{\mathrm{min}}=275.8$ & $0.480$ & $0.031$\\
 & $153.6$ & $313.4$ &  $\tau_{\mathrm{bb}}=575.9$ & $0.810$ & $0.058$\\
\midrule
SiO$_{2}$ & $106.6$ & $217.6$ &  $\tau_{\mathrm{min}}=6.6\times10^{3}$ & $0.051$ & $0.001$\\
 & $368.9$ & $960.0$ &  $\tau_{\mathrm{bb}}=5.0\times10^{4}$ & $0.136$ & $0.020$\\
\bottomrule
\end{tabular}
\label{table:table_tau_r_a_single}
\end{table}

\section*{Analyzing the acceleration time of core-(multi-)shell light sails}
Armed with the data for the aforementioned individual materials, we can now observe how they behave when combined to form core-(multi-)shell particles. We consider again a metasurface in the form of a square lattice with lattice constant $\Lambda$ comprising identical core-(multi-)shell spheres.

\begin{figure}[h!]
\centering
\includegraphics[width=1.0\textwidth]{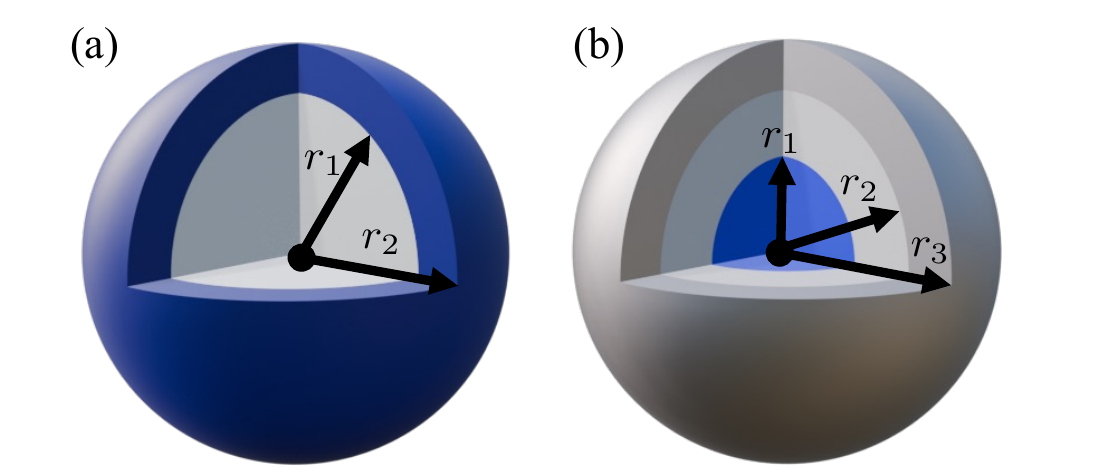}
\caption{A schematic of (a) a core-shell constituent sphere in the lattice with inner and outer radii $r_{1}$ and $r_{2}$ respectively. The same is given in (b) for a core-double-shell sphere with inner, central, and outer radii $r_{1}$, $r_{2}$ and $r_{3}$, respectively.}
\label{fig:layered_spheres}
\end{figure}

Here, we highlight a core-shell combination of Si and SiO$_{2}$ that yields a high broadband reflectance, where the inner and outer materials have radii $r_{1}$ and $r_{2}$, respectively (c.f. Fig.~\ref{fig:layered_spheres}~(a)). Additionally, out of the materials we consider, SiO$_{2}$ and Si have the lowest densities, which is conducive to a lower acceleration time. Such a structure provides an excellent example of when a combination of dielectric materials with significantly different permittivities can combine to produce higher reflectances than a metallic structure, a scenario that has also been documented in works done on Bragg reflectors~\cite{convertino1997organic, dai2016design, schubert2007distributed, kim2021broadband, palo2023prospects, zhang2019distributed}. Moreover, we will discuss core-double-shell structures containing combinations of Si, SiO$_{2}$, and Al, each containing an inner core and a central and an outer shell with radii $r_{1}$, $r_{2}$, and $r_{3}$, respectively (c.f. Fig.~\ref{fig:layered_spheres}~(b)). 

To locate the corresponding optimal geometrical parameters, we sweep across values for the lattice constant and radii, ensuring that $r_{1}<r_{2}<\Lambda/2$ for the core-shell structure, and $r_{1}<r_{2}<r_{3}<\Lambda/2$ for the core-double-shell structures, where $\Lambda<\lambda_{0}$ at all times. For numerical efficiency, we express the outer radius as fractions of the lattice constant and the inner radius(radii) as fractions of the outer radius(radii) as follows: 
\begingroup 
\addtolength\jot{5pt}
    \begin{numcases}{\text{Core-shell:}}
        0.01\leq \frac{r_{2}}{\Lambda}\leq0.49,\label{ineq:2l_1}\\
         0.01\leq \frac{r_{1}}{r_{2}}\leq0.99.\label{ineq:2l_2}
    \end{numcases}
\endgroup    
\begingroup 
\addtolength\jot{5pt}
    \begin{numcases}{\text{Core-double-shell:}}
         0.01\leq \frac{r_{3}}{\Lambda}\leq0.49,\label{ineq:3l_1}\\
         0.01\leq \frac{r_{2}}{r_{3}}\leq0.99,\label{ineq:3l_2}\\
         0.01\leq \frac{r_{1}}{r_{2}}\leq0.99.\label{ineq:3l_3}
    \end{numcases} 
\endgroup

\begin{figure}[ht]
\centering
\includegraphics[width=1\textwidth]{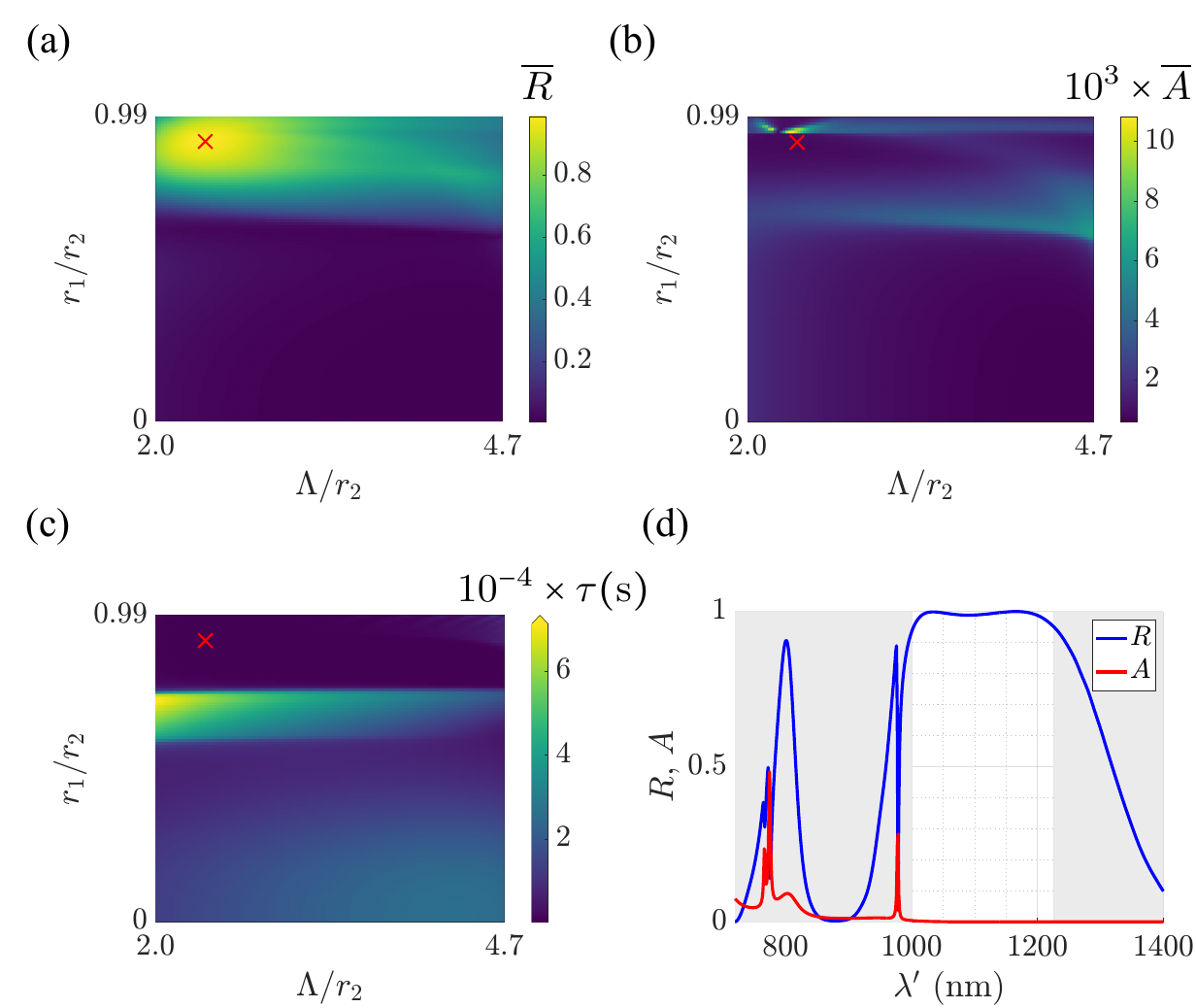}
\caption{In (a), (b), and (c), the average reflectance $\overline{R}$, average absorptance $\overline{A}$, and acceleration time $\tau$, respectively are plotted as a functions of $r_{1}/r_{2}$ and $\Lambda/r_{2}$ for $r_{2}=\SI{204.3}{\nano\meter}$ and lattice constant $\Lambda$ for a light sail containing Si/SiO$_{2}$ spheres. The red cross in all three plots signifies the location of the maximum average reflectance $\overline{R}_{\mathrm{max}}=0.988$, $\overline{A}=9.2\times10^{-4}$ and $\tau_{\mathrm{bb}}=\SI{395.3}{\second}$ given in Table~\ref{table:table_tau_r_a}. In (d) are the reflectance $R$ and absorptance $A$, which correspond to $\overline{R}_{\mathrm{max}}$. The range of interest $\lambda'\in[1.0\lambda_{0}, 1.225\lambda_{0}]$ is highlighted between the two gray regions.}
\label{fig:ref_abs_best_colour}
\end{figure}

As done in the previous section, we consider the average reflectance $\overline{R}$ to 
locate material parameters that give rise to a high broadband reflectance and low broadband absorptance. In doing this, one finds that the maximum average reflectance for the Si/SiO$_{2}$ combination has a value of $\overline{R}_{\mathrm{max}}=0.988$ with geometrical parameter values given by $r_{1}=\SI{186.2}{\nano\meter}$, $r_{2}=\SI{204.3}{\nano\meter}$ and $\Lambda=\SI{494.3}{\nano\meter}$ (c.f. Table~\ref{table:table_tau_r_a}). For $r_{2}=\SI{204.3}{\nano\meter}$, we present in Figs.~\ref{fig:ref_abs_best_colour}~(a),~(b),~and~(c) $\overline{R}$, $\overline{A}$, and $\tau$ as a function of $r_{1}/r_{2}$ and $\Lambda/r_{2}$, where $r_{1}$ obeys inequality~\eqref{ineq:2l_2}. The red cross in Figs.~\ref{fig:ref_abs_best_colour}~(a),~(b)~and~(c) depicts the locations of $\overline{R}$, $\overline{A}$, and $\tau$ corresponding to $\overline{R}_{\mathrm{max}}$. Plots of the reflectivity $R(\lambda')$ and absorptance $A(\lambda')$ corresponding to $\overline{R}_{\mathrm{max}}$ are given in Fig.~\ref{fig:ref_abs_best_colour}~(d). 

Note that, despite the Si/SiO$_{2}$ combination producing a high broadband reflectance, the acceleration time $\tau_{\mathrm{bb}}$ is $\SI{68.6}{\second}$ longer than that of a light sail made of pure silicon spheres, which also demonstrates a high broadband reflectance (c.f. Tables~\ref{table:table_tau_r_a_single} and~\ref{table:table_tau_r_a}). The increased acceleration time of the Si/SiO$_{2}$ combination is due to the larger total radius and mass of each sphere compared to pure silicon. However, an important factor to consider is the temperature resistance of the light sail. As mentioned in Refs.~\cite{holdman2022thermal, ilic2018nanophotonic}, the inclusion of SiO$_{2}$ can lead to increased radiative cooling, reducing the risk of heat damage compared to the light sail consisting of pure silicon.

In Table~\ref{table:table_tau_r_a}, we also provide data for other considered material combinations. We emphasize again why one may wish to consider $\tau_{\mathrm{bb}}$ from a high broadband reflectance and low broadband absorptance as opposed to a numerically minimized acceleration time $\tau_{\mathrm{min}}$, where no previous assumptions are made about the reflectance or absorptance. As seen for the SiO$_{2}$/Si/Al combination in Table~\ref{table:table_tau_r_a}, the average absorptance takes a large value ($\overline{A}=0.449$). Still, the numerically minimized acceleration time is very low ($\tau_{\mathrm{min}}=\SI{102.2}{\second}$). However, Eq.~\eqref{eq:acc_time} does not depend on temperature, and such a high absorptance could realistically lead to thermal deformation. As a result, ensuring a high broadband reflectance and low broadband absorptance could be preferred.


\begin{table}[h!]
\bigskip
\centering
\caption{The numerically minimized acceleration times $\tau_{\mathrm{min}}$ and those corresponding to maximum broadband reflectance ($\tau_{\mathrm{bb}}$), along with the average reflectance $\overline{R}$, average absorptance $\overline{A}$, radii $r_1$, $r_2$, and $r_3$ (for spheres with two shells), and lattice constants $\Lambda$ for lattices of spheres made from Si and SiO$_{2}$, and four combinations containing some or all the materials Al, Si, and SiO$_{2}$.}
\begin{tabular}{rrrrrrrr}
\toprule
Materials & $r_{1}$ (nm) & $r_{2}$ (nm) & $r_{3}$ (nm) & $\Lambda$ (nm) &$\tau$ (s) & $\overline{R}$ & $\overline{A}$\\
\midrule
Si/SiO$_{2}$ & $163.6$ & $172.1$ & $-$ & $941.4$ & $\tau_{\mathrm{min}}=211.9$ & $0.455$ & $9.6\times10^{-4}$\\
 & $186.2$ & $204.3$ & $-$ & $494.3$ & $\tau_{\mathrm{bb}}=395.3$ & $0.988$ & $9.2\times10^{-4}$\\
\midrule
SiO$_{2}$/Si/SiO$_{2}$ & $122.8$ & $176.4$ & $253.4$ & $941.4$ & $\tau_{\mathrm{min}}=400.5$ & $0.575$ & $0.005$\\
 & $147.4$ & $206.0$ & $287.9$ & $587.5$ & $\tau_{\mathrm{bb}}=791.4$ & $0.938$ & $0.005$\\
\midrule
Al/SiO$_{2}$/Si & $89.3$ & $132.1$ & $195.3$ & $922.7$ & $\tau_{\mathrm{min}}=213.8$ & $0.494$ & $0.120$\\
 & $4.1$ & $27.6$ & $187.4$ & $475.7$ & $\tau_{\mathrm{bb}}=335.8$ & $0.991$ & $4.4\times10^{-4}$\\
\midrule
Si/Al/SiO$_{2}$ & $11.7$ & $12.8$ & $14.0$ & $28.6$ & $\tau_{\mathrm{min}}=210.1$ & $0.077$ & $0.329$\\
 & $200.2$ & $202.2$ & $204.3$ & $494.3$ & $\tau_{\mathrm{bb}}=435.0$ & $0.848$ & $0.131$\\
\midrule
SiO$_{2}$/Si/Al & $12.4$ & $13.1$ & $13.8$ & $28.6$ &$\tau_{\mathrm{min}}=102.2$& $0.224$ & $0.449$\\
 & $0.02$ & $1.5$ & $150.9$ & $308.0$ & $\tau_{\mathrm{bb}}=566.6$ & $0.810$ & $0.057$\\
\bottomrule
\end{tabular}
\label{table:table_tau_r_a}
\end{table}

\section*{Analyzing the high broadband reflectance of a core-shell Si/SiO$_{2}$ light sail}
To see why we obtain such a high broadband reflection for the Si/SiO$_{2}$ combination in our desired wavelength range, it makes sense to consider the effect of lattice interactions. To do this, we utilize Eq.~(23a) from Ref.~\cite{rahimzadegan2022comprehensive}, which expresses the transmission $t$ as a function of the multipolar interactions. Specifically, we have 
\begin{align}\label{eq:transmission}
    t 
    &=
    1 - (3\widetilde{a_{1, \mathrm{eff}}}+3\widetilde{b_{1, \mathrm{eff}}} + 5\widetilde{a_{2, \mathrm{eff}}} + 5\widetilde{b_{2, \mathrm{eff}}})\, ,
\end{align}
where $\widetilde{a_{1, \mathrm{eff}}}$, $\widetilde{b_{1, \mathrm{eff}}}$, $\widetilde{a_{2, \mathrm{eff}}}$, and $\widetilde{b_{2, \mathrm{eff}}}$ are the normalized effective electric dipole, magnetic dipole, electric quadrupole, and magnetic quadrupole Mie coefficients, respectively. The tilde above each quantity represents a normalization concerning the normalized lattice constant $\widetilde{\Lambda}=\Lambda/\lambda'$, where
\begin{align}\label{eq:normalisation}
    \widetilde{a_{1, \mathrm{eff}}}
    &=
    \frac{a_{1, \mathrm{eff}}}{4\pi\widetilde{\Lambda}^{2}},\quad\mathrm{etc.}
\end{align}

Equation~(19) in Ref~\cite{rahimzadegan2022comprehensive} defines the effective Mie coefficients $a_{1, \mathrm{eff}}$, etc. Note that a similar analysis was done up to dipolar order by considering polarizabilities~\cite{evlyukhin2020lightweight}. We extend this here by including the effects of quadrupoles. We emphasize that an analysis up to quadrupolar order is fully sufficient since our data converges in our desired wavelength range with this approximation. Additionally, we consider the transmission coefficient $t$ instead of the reflection coefficient defined in~\cite[Eq.~(23b)]{rahimzadegan2022comprehensive} since the transmission coefficient reveals more information about the relationship between the real and imaginary parts of each of the multipoles, highlighting more clearly the effects of lattice interactions.

\begin{figure}[h!]
\centering
\includegraphics[width=1\textwidth]{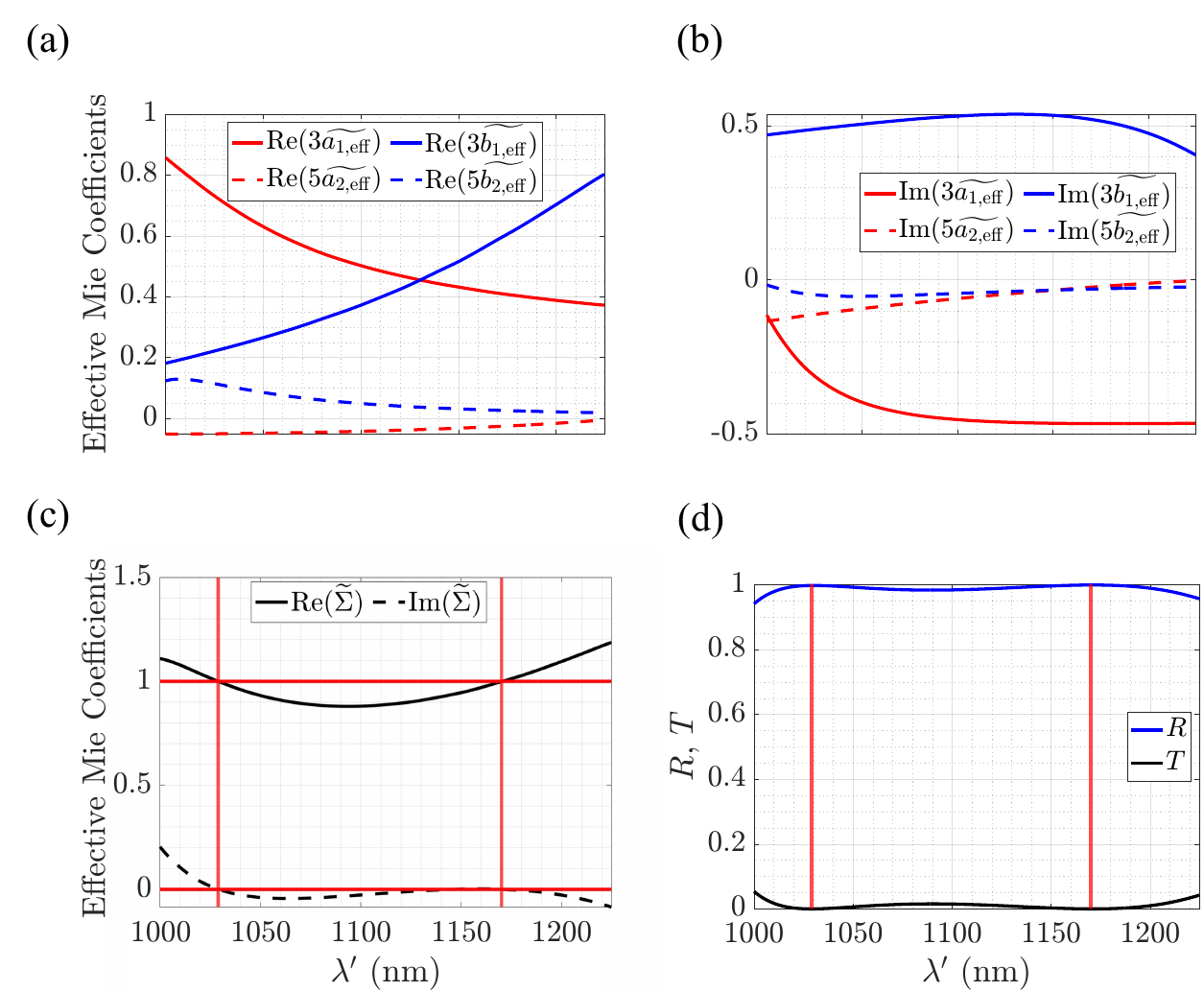}
\caption{In (a) and (b), the real and imaginary parts of the normalized electric and magnetic dipolar and quadrupolar Mie coefficients (with the prefactors present in Eq.~\eqref{eq:sum_tilde}) are respectively given as a function of the Doppler shifted wavelength $\lambda'$. The accompanying sums of the data in (a) and (b) are given in (c), along with red vertical and horizontal lines, which at their intersection points correspond to the combinations of multipoles that yield maximum reflectance. In (d) are the corresponding reflectance and transmittance spectra, with vertical red lines showing the maximum reflectance values.}
\label{fig:coeffs_ref_tra}
\end{figure}

Relabeling Eq.~\eqref{eq:transmission} as
\begin{align}\label{eq:sum_relabelled}
    t 
    &=
    1 - \widetilde{\Sigma}\, ,
\end{align}
where 
\begin{align}\label{eq:sum_tilde}
\widetilde{\Sigma} 
&=
3\widetilde{a_{1, \mathrm{eff}}}+3\widetilde{b_{1, \mathrm{eff}}} + 5\widetilde{a_{2, \mathrm{eff}}} + 5\widetilde{b_{2, \mathrm{eff}}}\, , 
\end{align}
we see that the most optimal configuration occurs when $\mathrm{Re}(\widetilde{\Sigma})= 1$ and $\mathrm{Im}(\widetilde{\Sigma})=0$, since this corresponds to zero transmission and, as the absorption always remains very close to zero (cf. Fig~\ref{fig:ref_abs_best_colour}~(c)), approximately total reflection. Plots of the real and imaginary parts of each normalized effective Mie coefficient (with the prefactors present in Eq.~\eqref{eq:sum_tilde}) that correspond to the maximum average reflectance $\overline{R}_{\mathrm{max}}$, as well as $\mathrm{Re}(\widetilde{\Sigma})$ and $\mathrm{Im}(\widetilde{\Sigma})$ are given in Figs.~\ref{fig:coeffs_ref_tra}(a),~(b), and~(c), respectively. The intersecting horizontal and vertical red lines in Fig.~\ref{fig:coeffs_ref_tra}(c) show the values for the sums of the real and imaginary parts of the effective Mie coefficients which correspond to maximum reflectance, that is, where $\mathrm{Re}(\widetilde{\Sigma})\approx1$ and $\mathrm{Im}(\widetilde{\Sigma})\approx0$. One sees in Fig.~\ref{fig:coeffs_ref_tra}(c) that Re$(\widetilde{\Sigma})$ and Im$(\widetilde{\Sigma})$ remain close to 1 and 0 over all wavelengths, respectively, which explains the high broadband reflectance $R(\lambda')$ shown in Fig.~\ref{fig:coeffs_ref_tra}(d), where the transmittance $T(\lambda')$ is also included. The vertical red lines in Fig.~\ref{fig:coeffs_ref_tra}(d) correspond to the points of maximum reflectance. 

\begin{figure}[h!]
\centering
\includegraphics[width=1\textwidth]{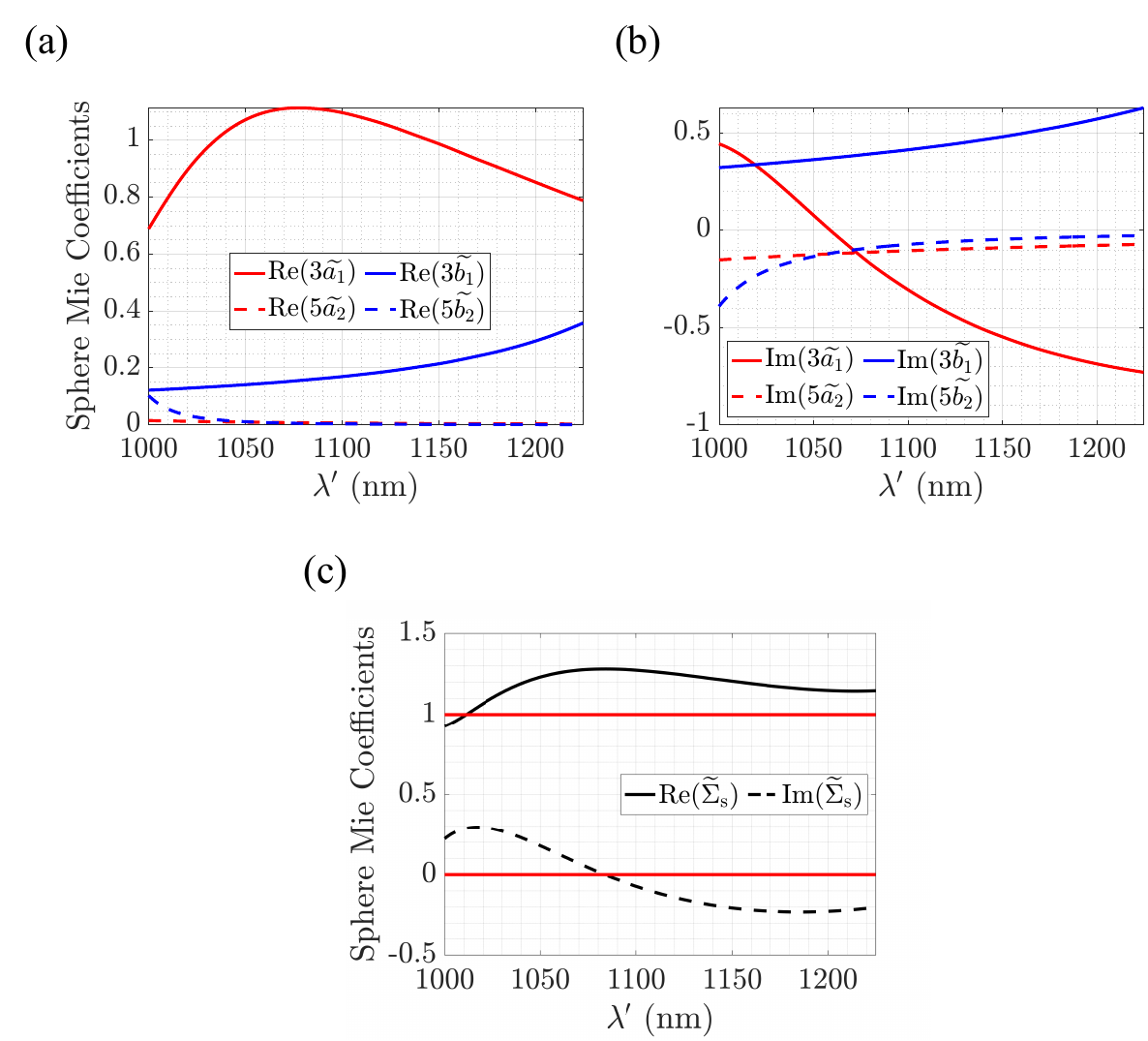}
\caption{The analogous plots to those in Figs.~\ref{fig:coeffs_ref_tra}~(a), (b), and (c) but for the Mie coefficients of a single sphere. The `s' subscript in (c) denotes `sphere'.}
\label{fig:coeffs_single_sphere}
\end{figure}

Comparing Figs~\ref{fig:coeffs_ref_tra}(a), (b), and (c) with the corresponding plots for the normalized Mie coefficients of a single sphere $\widetilde{a_{1}}, \widetilde{b_{1}}, \widetilde{a_{2}}$, and $\widetilde{b_{2}}$, and $\widetilde{\Sigma}_{\mathrm{s}}$ (where the `s' subscript stands for `sphere') in Figs~\ref{fig:coeffs_single_sphere}(a), (b), and (c), one notices the effect of the lattice interactions on the distributions of the individual multipolar responses. Note that we use the same normalization for a single sphere as that for the lattice in Eq.~\eqref{eq:normalisation} for consistency.  

\section*{The effects of an embedding surrounding the Si/SiO$_{2}$ lattice}

\begin{figure}[h!]
\centering
\includegraphics[width=1.0\textwidth]{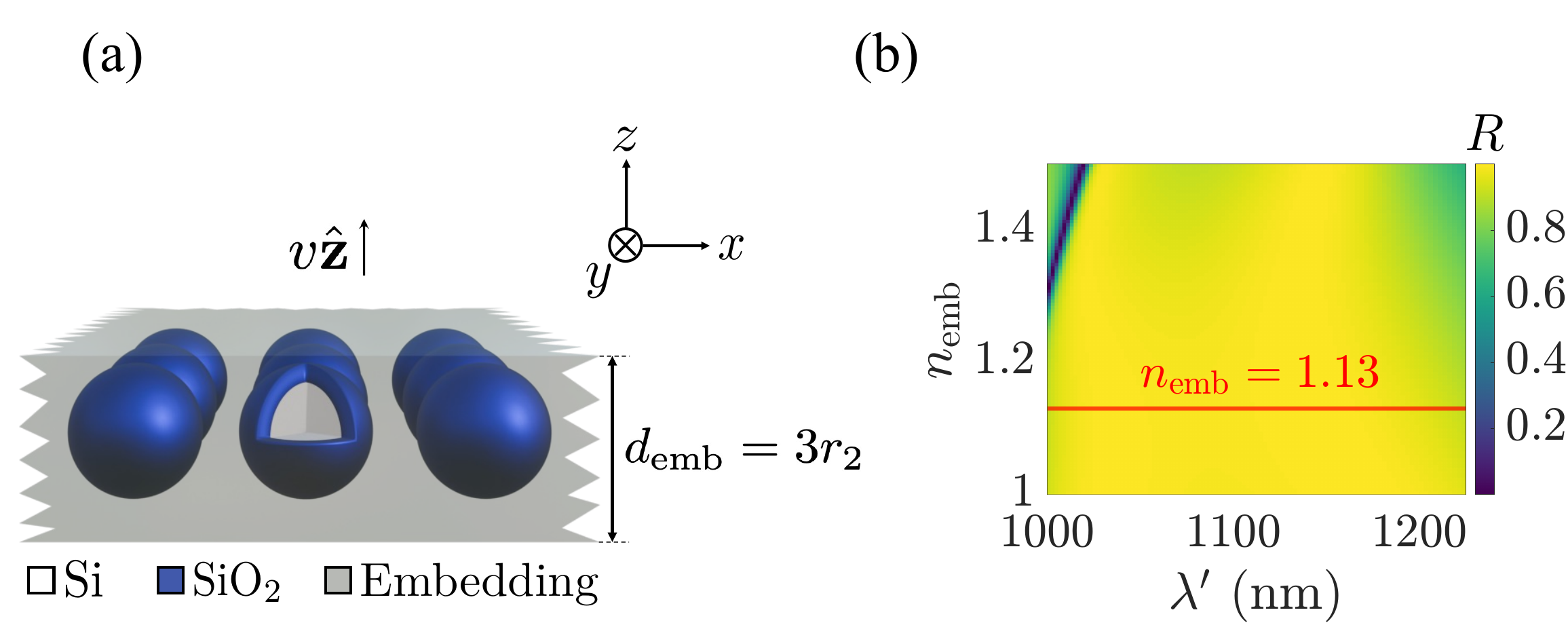}
\caption{In (a) is the schematic of a core-shell Si/SiO$_{2}$ spherical array moving with speed $v$ along the $+z$--axis embedded in a supporting medium with thickness $d_\mathrm{emb}=3r_{2}$, where $r_{2}$ is the radius of each shell. The color plot in (b) shows the reflectance $R$ for the optimized Si/SiO$_{2}$ array as a function of the embedding refractive index $n_\mathrm{emb}$ and Doppler shifted wavelength $\lambda'$, along with the refractive index $n_\mathrm{emb}=1.13$ above which the minimum reflectance drops below 90\% (red horizontal line).}
\label{fig:varying_embedding}
\end{figure}

Now that we have optimized a system containing core-shell spheres in vacuum, we can study the effects of adding an embedding to support the Si/SiO$_{2}$ lattice. As an example, we consider an embedding with thickness $d_{\mathrm{emb}}=3r_{2}$ (c.f. Fig.~\ref{fig:varying_embedding}~(a)) and refractive index $n_{\mathrm{emb}}$ and plot the dependence of the reflectance on $\lambda'$ and $n_{\mathrm{emb}}$ in Fig.~\ref{fig:varying_embedding}~(b). If $n_{\mathrm{emb}}$ remains low enough, one would still achieve a minimum reflectance greater than $\SI{90}{\percent}$. With this condition in mind, the current parameter combination for Si/SiO$_{2}$ holds up to $n_{\mathrm{emb}}\approx1.13$ (c.f. Fig.~\ref{fig:varying_embedding}(b)). A refractive index $n_{\mathrm{emb}}>1.13$ would require finding new parameters. To optimize the acceleration time, however, one would have to be careful that the embedding is not too dense such that it adds excessive mass to the system. An example of a possible embedding material  is polydimethylsiloxane (PDMS) as considered in Refs.~\cite{evlyukhin2020lightweight, lien2022experimental}, which has a refractive index of $n_{\mathrm{emb}}=1.45$ and a very low density of $\rho_{\mathrm{PDMS}}=\SI{965}{\kilogram\meter}^{-3}$.

\section*{Conclusions}

In this work, we explored the reflectance $R$, absorptance $A$, and acceleration time $\tau$ of various lattices of homogeneous and core-(multi-)shell spheres using the re-normalized T--matrix obtained via Mie theory. The materials we considered were aluminum (Al), silicon (Si), and silicon dioxide (SiO$_2$).

Our primary goal was to design lattices that could be used as light sails in future outer space explorations, where the light sails are to be accelerated by an Earth-based laser array to 20\% of the speed of light to carry micro-satellites to neighboring solar systems. Due to Doppler shifts, the sail observes different wavelengths depending on its speed, thus requiring a high broadband reflectance for maximum momentum transfer from the lasers. We identified materials that provided a high broadband reflectance and low absorptance, where the latter is crucial to reducing unwanted heating and deformation. From this, we calculated the acceleration times of each light sail, incorporating the mass and spacing of the spheres.

We highlighted a lattice type that fulfills the conditions mentioned above, namely a lattice comprising core-shell spheres with a Si core and SiO$_{2}$ shell, yielding $\overline{R}_{\mathrm{max}} = 0.988$ and a corresponding acceleration time of $\tau = \SI{395.3}{\second}$. Furthermore, we noted the potential risk of minimizing the acceleration time numerically without any prior assumptions about the reflectance and absorptance since Eq.~\eqref{eq:acc_time} defining $\tau$ depends on the absorptance but not the resulting temperature changes, which could cause thermal deformation in the light sail. In other words, $\tau$ can be small, but the corresponding absorptance could be high.  

Next, the high broadband reflectance of the Si/SiO$_2$ combination was explained using equations from~\cite{rahimzadegan2022comprehensive} to assess the effect of lattice interactions on the multipolar components of the outgoing field up to quadrupolar order. Additionally, we examined the effect of adding embeddings of different refractive indices to the Si/SiO$_{2}$ lattice, finding that refractive indices up to $1.13$ still yield over 90\% reflectance without the need to search for new material parameters.

Our results contribute significantly to light sail physics by proposing structures with core-shell spheres, a previously unexplored area in this context. Thus, they open avenues for future experimental work.

\section*{Acknowledgements} M.R.W. and C.R. acknowledge support from the Max Planck School of Photonics, supported by BMBF, the Max Planck Society, and the Fraunhofer Society. M.R.W. and L.R. acknowledge support from the Karlsruhe School of Optics and Photonics (KSOP). L.R. acknowledges support by the Deutsche Forschungsgemeinschaft (DFG, German Research Foundation) - Project-ID 258734477 - SFB 1173. B.Z. and C.R. acknowledge support by the KIT through the ``Virtual Materials Design'' (VIRTMAT) project. 

\bibliography{references}

\end{document}